20/01/2009

# Power Spectrum of Out-of-equilibrium Forces in Living Cells : Amplitude and Frequency Dependence

François Gallet*, Delphine Arcizet[§], Pierre Bohec and Alain Richert

Laboratoire Matière et Systèmes Complexes

UMR 7057 associée au CNRS et à l'Université Paris7 - Denis Diderot

Bâtiment Condorcet - Case courrier 7056

Université Paris-Diderot

75205 Paris cedex 13 – France

* corresponding author : francois.gallet@univ-paris-diderot.fr

[§] present address : Center for Nanosciences (CeNS) and Faculty of Physics - Ludwig-Maximilians Universitaet - Geschwister-Scholl-Platz 1, D- 80539 Muenchen, Germany

**Summary**

Living cells exhibit an important out-of-equilibrium mechanical activity, mainly due to the forces generated by molecular motors. These motor proteins, acting individually or collectively on the cytoskeleton, contribute to the violation of the fluctuation-dissipation theorem in living systems. In this work we probe the cytoskeletal out-of-equilibrium dynamics by performing simultaneous active and passive microrheology experiments, using the same micron-sized probe specifically bound to the actin cortex. The free motion of the probe exhibits a constrained, subdiffusive behavior at short time scales (t < 2s), and a directed, superdiffusive behavior at larger time scales, while, in response to a step force, its creep function presents the usual weak power law dependence with time. Combining the results of both experiments, we precisely measure for the first time the power spectrum of the force fluctuations exerted on this probe, which lies more than one order of magnitude above the spectrum expected at equilibrium, and greatly depends on frequency. We retrieve an effective temperature $T_{eff}$ of the system, as an estimate of the departure from thermal equilibrium. This departure is especially pronounced on long time scales, where $T_{eff}$ bears the footprint of the cooperative activity of motors pulling on the actin network. ATP depletion reduces the fluctuating force amplitude and results in a sharp decrease of $T_{eff}$ towards equilibrium.

20/01/2009

**Introduction**

Living cells are constitutively out-of-equilibrium systems. Indeed, biological activity is characterized by constant exchanges of matter and energy with the environment, and biochemical reactions provide the power supply necessary to the metabolism. Non-equilibrium activity allows cell motion and deformation when needed for function or survival, and cargo transport within the dense microenvironment of the cell. These mechanical effects are directly related to the dynamics of polymerization/depolymerization of the cytoskeleton and to the activity of molecular motors, which convert the chemical energy stored in ATP into mechanical energy.

Consequently, the fluctuation-dissipation theorem is violated within a living cell. The cell viscoelastic properties cannot be simply related to thermal fluctuations, but are expected to be mainly controlled by the active forces generated along the cytoskeleton. In the past years, both active and passive microrheology have been used to characterize the mechanical and rheological properties within specific regions of living cells, and their implications in essential biological functions like cell adhesion and migration, or cargo transport inside the cytoplasm. These two techniques do not provide the same information.

In active microrheology, the complex viscoelastic modulus $G(\omega) = G'(\omega)+iG''(\omega)$ of the intracellular medium is determined by applying an oscillating force to a micrometric probe bound to the cytoskeleton. Experiments are driven either in the cell cortex, by using a bead specifically bound to transmembrane receptors [1-3], or in the bulk cellular body [4, 5], using magnetic beads embedded in the cytoplasm. Alternatively, the creep function $J(t)$ characterizes the cell response to a step of force of given amplitude [6-8]. Active rheology is an intrusive technique, but presents the advantage to give direct access to the elastic and dissipative parts of the mechanical response. Most of the works report a power law dependence of the complex viscoelastic modulus with frequency, or of the creep function with time [9, 10]. This behavior, similar to the one of crosslinked polymer networks, and of soft glassy materials, indicates that the dissipation time scales are widely and densely distributed in the system. Although it has been shown that the activity of motor proteins plays a determinant role in this mechanical response, there is yet no global model accounting for the coupling between biological activity and mechanical behavior.

By contrast, passive microrheology allows to probe different regions of the cell in a non-perturbing manner, using either endogenous granules [11] or submicron-sized particles embedded within the cell body [12-15]. Analogous experiments were performed close to the membrane, using particles specifically attached to it [6, 16-19] or vesicles moving in the cortical region [20]. The mean square displacement of the particle exhibits different regimes, depending on whether the particle is constrained by the fluctuations of the network, or linked to molecular motors which impart a directed motion to it. However, it is generally not possible to deduce the rheological properties of the material from such experiments, because the fluctuation-dissipation theorem does not apply in living systems. Retrieving the dissipative coefficients from the free diffusive motion of the particle would require knowing the distribution of fluctuating forces exerted on the particle. This force spectrum is related to many active physiological processes in living cell, and cannot be easily evaluated.

Recently, a combination of active and passive microrheology was performed in an *in vitro* actin gel crosslinked by myosin motors[21]. A departure from equilibrium was observed over long time scales, due to active contractions in the network. In living systems, the litterature concerning active and passive microrheology on the same system is scarce, and in these cases the validity of the fluctuation-dissipation theorem was not assessed[6, 19]. The first parallel analysis of the free translational motion of a magnetic phagosome embedded in an amoeba, and of the rotation of a phagosome chain in an external magnetic field, showed a departure from equilibrium, and allowed calculating an effective temperature in the intracellular medium[22].

This work presents the first measurements of both active and passive microrheology in a living mammalian cell, performed under strictly identical conditions with the same probe bound to the cortical actin network, in order to quantify the departure from thermal equilibrium. We use muscular cells, in which the activity of myosin molecular motors is expected to be particularly prominent. We successfully retrieve the mean square displacement of the free bead motion, and its creep function in response to an external force step applied with optical tweezers. From the two datasets we infer a numerical value for the power spectrum of active forces exerted on the bead, in the frequency range 0.01 – 100 Hz.

At high frequencies (s > 0.5Hz), the force power spectrum approximately follows the same frequency dependence as an equilibrium system, but at low frequencies (s < 0.5Hz), the force power spectrum behaves likes $s^{-1.74}$, corresponding to a directed motion of the probe under cooperative action of molecular motors. In both regimes, we retrieve an effective temperature $T_{eff}$ higher than the actual thermodynamic temperature (from 13 to 100 times larger). $T_{eff}$ can be considered as a scaling for the out-of-equilibrium, frequency-dependant, biological activity. ATP depletion drastically decreases the effective temperature, supporting evidence for the active biological origin of fluctuating forces.

**Formulation of the out-of-equilibrium Fluctuation-Dissipation Theorem.**

The motion of a micron-sized probe embedded into a viscoelastic medium is governed by the Langevin equation, which relates the probe velocity v= dx/dt to the total force exerted on it, sum of the externally applied force $F_{ext}$ and of the random internal fluctuating force $F_r$ [13]:

$$m\frac{dv}{dt} + m\int_{-\infty}^{+\infty}\gamma(t-t')v(t')dt' = F_{ext}(t) + F_r(t) \quad [1]$$

Here m is the probe mass, and γ(t) a delayed friction kernel describing the viscoelastic properties of the medium. In order to respect the causality principle, one has to assign γ(t)=0 for t<0. The validity of the Langevin equation does not depend on equilibrium or out-of-equilibrium conditions. It also holds for a probe bound to the medium, partially but not necessarily fully embedded into it, provided that the probe dynamics is controlled by the mechanical properties of this medium.

In a similar way, the creep function J(t) of the system can be defined as :

$$x(t) = \int_{-\infty}^{+\infty} J(t-t')dF(t') = \int_{-\infty}^{+\infty} J(t-t')\frac{dF}{dt'}\bigg|_{t'} dt' \quad [2]$$

J(t) represents the displacement of the probe generated by a force step of magnitude unity applied at t = 0 (as for γ(t), one has J(t)= 0 for t<0).

Introducing the autocorrelation functions $S_F(t) = <F_r(t')F_r(t'+t)>_{t'}$ and $S_v(t) = <v(t')v(t'+t)>_{t'}$, and taking the Laplace transform (LT) of equation [1], one can show that :

$$\hat{S}_v(s) = \left(\frac{1}{m(\hat{\gamma}(s)+s)}\right)^2 \hat{S}_F(s) = s^4(\hat{J}(s))^2 \hat{S}_F(s) \quad [3]$$

where $\hat{S}_v(s) = \int_0^\infty S_v(t)\exp(-st)dt$ and $\hat{S}_F(s) = \int_0^\infty S_F(t)\exp(-st)dt$ are respectively the power spectra of the probe velocity and of the random force in the Laplace frequency space.

Equation [3] establishes the key relation, valid in any conditions, between the spectrum of the free velocity fluctuations in the absence of external force $\hat{S}_v(s)$, the mechanical response to an externally applied force $\hat{J}(s)$, and the spectrum of fluctuating forces $\hat{S}_F(s)$ exerted on the probe.

In the particular case of thermal equilibrium, the usual fluctuation-dissipation theorem applies, and the force spectrum is determined by thermal fluctuations, according to [23]:

$$\hat{S}_{Feq}(s) = \frac{k_B T}{s^2 \hat{J}(s)} \qquad [4]$$

In such conditions, the fluctuation-dissipation theorem can be written under the form of the Einstein relation:

$$\hat{S}_{veq}(s) = \frac{k_B T}{m(\hat{\gamma}(s) + s)} \qquad [5]$$

Now, in the out-of-equilibrium case, the force fluctuation spectrum is not determined by the temperature, but equation [3] remains valid. Following several authors, it may be convenient to generalize equation [4] by introducing an effective temperature for the system, defined as:

$$\hat{S}_F(s) = \frac{k_B T_{eff}}{s^2 \hat{J}(s)} \qquad [6]$$

In other words, $\theta(s) = \frac{T_{eff}}{T} = \frac{\hat{S}_F(s)}{\hat{S}_{Feq}(s)}$ may be considered as an index of the departure from thermal equilibrium for the force power spectrum. Notice that $T_{eff}$ is not a real thermodynamical variable in the sense that it may depend on the frequency s.

Actually, for a freely diffusing probe, the mean square displacement (MSD) $\Delta x^2(t) = <(x(t')-x(t'+t))^2>_{t'}$ is a more accessible quantity than the velocity correlation function $S_v(t) = <v(t)v(t+t')>_{t'}$. Both are related through $s^2 \Delta \hat{x}^2(s) = 2\hat{S}_v(s)$ (the factor 2 comes from the definition of $\Delta x^2(t)$, which is not exactly an autocorrelation function). Consequently, one can rewrite equation [3] under a form more adapted to experimental analysis:

$$\hat{S}_F(s) = \frac{\Delta \hat{x}^2(s)}{2s^2 (\hat{J}(s))^2} \qquad [7]$$

Equation [7] will be used in the next section to determine the force fluctuation spectrum $\hat{S}_F(s)$, from the independent measurement of $\Delta \hat{x}^2(s)$ and $\hat{J}(s)$ by a passive and active microrheology experiment, performed quasi-simultaneously on the same probe in identical conditions.

In the following we will encounter the situation where both the MSD and the creep function behave as power laws of time: $\Delta x^2(t) \propto t^\beta$ and $J(t) \propto t^\alpha$. Then it is easy to show that the force power spectrum is expected to take the form[13]:

$$\hat{S}_F(s) \propto s^{2\alpha-\beta-1} \Leftrightarrow <F_r(t')F_r(t'+t)>_{t'} \propto t^{\beta-2\alpha} \qquad [8]$$

In the equilibrium case, one derives from equation [4] the additional relationship:

$$\hat{S}_{Feq}(s) \propto s^{\alpha-1} \Leftrightarrow <F_r(t')F_r(t'+t)>_{t'} \propto t^{-\alpha} \qquad [9]$$

**Experimental**

*Cell culture*

The myogenic cell line C2-C12 (kindly provided by M. Lambert and R. M. Mège, INSERM U440, Institut du Fer à Moulin, Paris) is a subclone of the C2 line derived from the skeletal muscle of adult CH3 mice [24]. Cells were grown at 37°C in a humidified 5% $CO_2$ - 95% air atmosphere, in DMEM supplemented with 10% fetal calf serum, 2 mM glutamine, 100 units/ml penicillin and 50 mg/ml streptomycin. They were detached from culture flasks 24 hours before experiments and plated at a density of about 300 cells/mm$^2$, in complete culture medium with serum, on glass coverslips coated with fibronectin (5 μg/mL for 3h at room temperature).

The probes, carboxylated silica microbeads (1.56 μm diameter, Bangs Laboratories Inc.), were coated with a peptide (Peptide2000, Integra Lifescience) containing the arginine-glycine-aspartic tripeptide (RGD) sequence in order to bind specifically to cortical actin, through transmembrane integrin receptors. Before use, coated beads were incubated in DMEM supplemented with 1% BSA for 30 minutes at 37°C. Beads were then added to the cells (~ 5μg of beads per coverslip) and further incubated for 20 min at 37°C. Unbound beads were washed away with medium. A closed experimental chamber was made by sealing the coverslip to a microscope slide, separated by a 100 μm plastic film spacer. The chamber was mounted on a piezoelectric stage (Polytec PI), allowing displacements in the range 0-100 Hz (maximum excursion 80 μm). The whole set up was placed on the plate of an inverted microscope (Leica DMIRB). All measurements were performed in a 37°C thermalisation box (Life Imaging System). The total time during which the cells were studied after sealing the chamber did not exceed one hour.

ATP depletion was achieved by incubating the cells during 20-30 min in a solution containing 6mM deoxyglucose + 10 mM NaN3, after binding the beads. Cells were then rinsed with serum deprived culture medium before sealing the chamber.

The averaged values reported in the results section refer to 60 measurements over cells in control conditions, and 23 measurements in ATP-depleted conditions.

*Passive microrheology*

The free diffusion of a bead bound to the cell was tracked with a quadrant photodiode located close to the image plane of the objective, and illuminated by the microscope bright field light (for a detailed description of the experimental set up, see ref (25)). Signal amplification allowed getting the x and y coordinates of the bead in the observation plane with a resolution better than 10nm. The bead was usually followed during 2min. Signal acquisition at 500Hz was achieved under LABVIEW® (National Instruments). For both coordinates x and y, the MSD $\Delta x^2(t) = <(x(t')-x(t'+t))^2>_{t'}$ and its Laplace Transform $\Delta \hat{x}^2(s)$ were numerically calculated under a home made Fortran program$^§$.

*Active microrheology*

Each active microrheology experiment was performed on the same bead immediately after its free diffusion recording, leaving not enough time between the two experiments for any remodeling of the cytoskeletal network. A force step F, either along the x or y axis, was applied to the bead at t = 0 with optical tweezers. In order to maintain the amplitude F constant, the bead tracked by the quadrant photodiode was kept at a fixed position with respect to the trap, by means of a feedback loop acting on the piezoelectric stage [25]. The relative bead-cell displacement x(t), equal to the stage displacement, was recorded for further analysis. The force was applied during at most 60s, with an amplitude between 5 and 50 pN. In such conditions the bead displacement fell in the range 0.1 - 0.3μm, yielding a small enough cell deformation to stay in the linear deformation regime. The creep function $J_x(t)$ (resp. $J_y(t)$) was defined as the ratio x(t)/F (resp. y(t)/F). The Laplace transforms $\hat{J}_x(s)$ and $\hat{J}_y(s)$ were numerically calculated as described above. We are aware that, due to the partial embedding of the bead into the cell, the retrieved x(t) results of a combination of both bead translation and bead rotation. This may induce a slight overestimate of J(t), but does not change the general conclusions of our study.

20/01/2009

## Results

*Mean square displacement*

A typical mean square displacement curve $\Delta x^2(t)$ is plotted in Fig. 1.a., in the range 0.02-120s, for a 1.56μm silica bead specifically bound to the membrane of a C2-C12 cell, and linked to the actin cortical network. Together is shown the reference noise level (Fig. 1.c), recorded from one of the residual beads unbound to a cell but firmly attached to the glass coverslip. The MSD curve exhibits two regimes of diffusion: a subdiffusive regime at short time scales (t<3s), and a superdiffusive one at large time scales (t>3s). In each regime, $\Delta x^2(t)$ is proportional to $t^\beta$, here with $\beta_1 = 0.11$ and $\beta_2 = 1.68$, respectively in the sub- and super-diffusive case. A similar behavior is observed for all the studied cells, with $<\beta_1> = 0.120 \pm 0.008$ and $<\beta_2> = 1.49 \pm 0.06$, respectively in the range $0.02 < t < 0.5s$ and $t > 10s$. The sub- and super-diffusive regimes respectively correspond to a constrained motion of the bead at short time scales, combined to a directed drift motion at longer time scale, as clearly visible on the x-y trajectory. This is consistent with other reports concerning freely diffusing particles, either embedded within the cell, or bound to its membrane [6, 9, 13, 15-18]. The crossover between the two regimes occurs on average at $t \approx 2s$.

ATP depletion noticeably modifies the free diffusion behavior, as illustrated in Fig. 1.b. First, the global amplitude of the mean square displacement is reduced as compared to control conditions, by a factor of 1.5 to 10, which is not uniform over the whole time range. The subdiffusive behavior is still observable at short time scales, but about half of the cells do not exhibit superdiffusion at large time scales anymore. This is a strong evidence that, in control conditions, the diffusive motion is largely due to non-equilibrium forces generated by the biological activity. On average, both exponents $\beta_1$ and $\beta_2$ are reduced, becoming: $<\beta_1> = 0.063 \pm 0.011$ and $<\beta_2> = 1.27 \pm 0.13$. Whenever it can be measured, the crossover time between the subdiffusive and superdiffusive regime is roughly unchanged, around 3s.

*Creep function*

Fig. 2 shows an example of creep response for a bead bound to a cell membrane and submitted to a force step. Like in most of the experiments reported in the literature [7, 8], the creep function $J(t) = x(t)/F$ exhibits a power law behavior over the full time range 0.02 - 30s. On Fig. 2, the best power law fit $J(t) = A(t/t_0)^\alpha$ yields $\alpha = 0.25$. Over 60 measurements, $\alpha$ is distributed in the range 0.05-0.4, and its average value is $<\alpha> = 0.183 \pm 0.015$. This is consistent with previous experiments on the same cell line: for cells submitted to uniaxial stretching, J(t) was shown to be accurately described by a power law, with an average value of 0.24 for the exponent α [7]. The origin of such power law behavior, thoroughly discussed elsewhere [6, 10], is related to the wide and dense distribution of response times in the system, spreading over several orders of magnitude.

For ATP-depleted cells, we observe a similar power behavior for the creep function, but the exponent α is noticeably smaller: $<\alpha>$ = 0.139 ± 0.015. Considering that α reflects the more elastic-like or dissipative-like character of the medium (see ref (3, 10)), this value, smaller than in control conditions, means that ATP depletion reduces the efficiency of the dissipative processes, and brings the cell closer to a rigid elastic solid. This is consistent with the picture that ATP depletion inhibits molecular motors activity and make the myosin heads bind to the actin filaments (rigor state), thus increasing the number of rigid crosslinks in the network.

*Force power spectrum*

The power spectrum of fluctuating forces $\hat{S}_F(s)$ exerted on the probe is inferred from the combination of passive and active microrheology experiments performed on the same probe (see eq. [7]). The mean spectrum $\hat{S}_F(s)$, geometrically averaged over 60 measurements for cells in control conditions, is plotted on Fig. 3. Together is also plotted the mean force spectrum $\hat{S}_{Feq}(s)$, calculated according to eq. [4] as if the system was in equilibrium. The actual power spectrum is always above the equilibrium one, over the full frequency range s = 0.05-50 Hz. This demonstrates the active origin of biological forces transmitted to the bead. A closer look at the active spectrum indicates the coexistence of two distinct power law behaviors $\hat{S}_F(s) \propto s^{-\gamma}$ according to the frequency. At high frequencies s>0.5 Hz (*i.e.* short time scale) the exponent value $\gamma_2$ = -0.81 is in agreement with the expected one $\gamma = 2<\alpha>-<\beta_1>-1$ = -0.75 (see equation [8]). Interestingly, one notices that $\gamma_2$ is also close to the exponent $\gamma_e$ = -0.80 measured on the equilibrium spectrum $\hat{S}_{Feq}(s)$. Thus, the actual force spectrum in this frequency range exhibits almost the same frequency dependence as the equilibrium one, although the prefactor is different: the ratio $\hat{S}_F(s)/\hat{S}_{Feq}(s)$ is here about 13. At lower frequencies s<0.5 Hz (i.e. larger time scales), the exponent of the power spectrum is shifted to $\gamma_1$ = -1.74, and the ratio $\hat{S}_F(s)/\hat{S}_{Feq}(s)$ rapidly increases with decreasing s. In this range the departure from equilibrium increases with time. The crossover at s ≈ 0.5 Hz (t ≈ 2s) corresponds to the transition between the subdiffusive and the superdiffusive regimes observed during free particle motion.

Another way to prove out-of-equilibrium conditions consists in calculating the effective temperature for the system (see equation [6]). The average index $\theta(s) = T_{eff}/T$ which quantifies the departure from equilibrium is plotted as a function of the frequency in figure 4 for control and ATP-depleted cells. In control conditions, $\theta(s)$ remains constant, close to 13, for $s > 0.5$ Hz, and rapidly increases as $s$ decreases below 0.5 Hz, which is consistent with figure 3. This approach using the effective temperature is quite adapted to show up the prominent effect of ATP depletion. As compared to control conditions, $\theta(s)$ is substantially lowered for ATP-deprived cells, indicating that ATP depletion considerably reduces the active part of force fluctuations and brings back the cells closer to equilibrium conditions. At high frequency, $\theta(s)$ is also constant, with a mean value of about 6 instead of 13 for control cells ; at lower frequency, $\theta(s)$ is reduced by a factor up to 10 and does not show a large increase with decreasing $s$ anymore. Although noticeably smaller than in the presence of ATP, $\theta(s)$ never reaches the value $\theta(s) = 1$ expected at thermodynamical equilibrium.

**Discussion**

By performing active and passive microrheology on the same micron-sized probe bound to the cell membrane, we were able to infer the precise amplitude and frequency dependence of the fluctuating forces spectrum exerted on this probe, bringing evidence that these forces are generated by active, out-of-equilibrium, biological processes. Several active mechanisms are likely to contribute to the force generation, and, at this stage, it is not trivial to unambiguously determine which one dominates. Indeed, the bead is linked to the cortical actin network through transmembrane receptors, and both active processes in the actin cortex and at the bead-cell contact may contribute to generate forces on the bead. Clearly, the free diffusion of the bead can be affected by local remodeling of the bonds in the contact area. However, it is reasonable to assume that the motion of the bead is mainly due to the activity of molecular motors, namely myosins, pulling on these filaments. This is supported by the observation that the typical time scale for contact remodeling is larger than 1min, which is beyond our experimental time range[19, 25]. Also, the force spectrum at low frequency exactly matches the superdiffusive regime of the particle motion, and this directed motion is known to be generated by myosin activity, which imprints a retrograde flow to actin filaments. Finally, preliminary experiments performed on cells treated with blebbistatin show that the force power spectrum is lowered to a level comparable to ATP-depleted cells when the myosin II activity is inhibited (data not shown). But we cannot yet completely rule out the possibility that other molecular motors partially contribute to the force spectrum. The results of previous works[14, 22] suggested that a large part of the active fluctuations in the bulk intracellular medium was due to active gliding along the microtubule network. However, our experiment is probably less sensitive to such processes, because it is performed in the cortical region, and also because the probes are not directly linked to the microtubule network.

As suggested in a recent theoretical work[26], the mechanism responsible for the bead motion can be depicted as follows: on short enough time scales, random binding and unbinding of individual myosin heads generates random forces on the filaments, while at longer time scales cooperative effects dominates to pull on the network and imprint a directed and centripetal motion to the bead. In the higher frequency regime s> 0.5 Hz, corresponding to shorter time scales t<2s, we observe that the spectrum has nearly the same frequency dependence as the equilibrium one, but its amplitude is larger by more than one order of magnitude. More precisely, the spectrum in this range can be approximated by $\hat{S}_F(s) \approx 9.1$ $s^{-0.81}$ pN².s. By taking the inverse Laplace transform, this leads to the expression of the force autocorrelation function $<F_r(t')F_r(t'+t)>_{t'} \approx 7.9\, t^{-0.19}$ pN². One notices that the amplitude of this function, a few square piconewtons, is comparable to the one of a single myosin motor. This supports the idea that, in this time range, the bead fluctuations are generated by molecular motors working independently. On the contrary, at lower frequencies, the spectrum takes the form $\hat{S}_F(s) \approx 4.1\, s^{-1.74}$ pN².s. The exponent $\gamma_1 = -1.74$ is reasonably close to the value $\gamma = -2$ suggested for the first time by Lau et al. for the spectrum frequency dependence [13]. In the real time space, this corresponds to $<F_r(t')F_r(t'+t)>_{t'} \approx 4.5\, t^{0.74}$ pN². The force-force autocorrelation function thus increases almost linearly with elapsed time, which can be explained only if the mean force exerted on the bead does not average to zero. This is consistent with the picture that, over long time, the bead is driven by a centripetal force generating its directed motion. Incidentally, we recall that the amplitude of force power spectrum is always larger than the equilibrium one, over the full explored time range. This differs from the observations reported in active in vitro gels [21], and in TC7 epithelial cells [9], where the force spectrum was found identical to the equilibrium one at short time scale.

The effective temperature index $\theta = T_{eff}/T$ is a convenient parameter to visualize the departure from equilibrium, although its interpretation as a true thermodynamical variable is not a trivial problem[27, 28]. On figure 4, the two plots of $\theta(s)$ enlighten the contribution of active processes in force generation, and make the comparison easier between normal and ATP-depleted cells. Interestingly, the important decrease of $\theta(s)$ in ATP-depleted conditions does not appear homogeneous over the full frequency range. In particular, the disappearance of the sharp increase of $\theta(s)$ at low frequencies is correlated to the absence of directed motion observed for a large fraction of the beads at large time: on average, the drift velocity of the bead is reduced from $0.22 \pm 0.03$ μm/min in normal conditions to $0.12 \pm 0.02$ μm/min in ATP-depleted conditions.

Surprisingly, for ATP-depleted cells, the effective temperature index $\theta(s)$ is reduced but never reaches the equilibrium value $\theta(s) = 1$, while one would expect ATP depletion to block all active processes, and to lead back to equilibrium conditions. Two reasons can be invoked to interpret this: i) ATP depletion may not be completely achieved in our experimental conditions. This is supported by the observation of a retrograde motion for a fraction of treated cells ; ii) other active processes, for instance microtubule dynamics which is GTP-dependant, may not be affected, or only partially affected by ATP depletion.

In conclusion, we report here a direct and quantitative characterization of the out-of-equilibrium mechanical activity in a living cell. Combining active and passive microrheology experiments performed at the same time, with the same probe, allowed retrieving the power spectrum of fluctuating forces generated by biological activity, which does not verify the usual fluctuation-dissipation theorem. We were able to precisely measure the amplitude and the time dependence of the force autocorrelation function, and to relate them to the forces generated by molecular motors, acting either individually at short time scale, or cooperatively at larger one. We quantitatively measured how ATP depletion induces a spectacular decrease of the force power spectrum towards its equilibrium value. Specific inhibition of myosin II activity is expected to confirm the dominant contribution of these molecular motors to the active processes in the cortical region. A better link between theoretical models, *in vitro* studies using active polymer networks, and *in vivo* experiments is still needed, but this study is an important step in this direction: it describes precisely and quantitatively the out-of-equilibrium force generation that governs cell mechanical behavior.

**Ackowledgements**

The authors acknowledge Noëlle Pottier, Claire Wilhelm, Damien Robert and Jean-Pierre Henry for interesting discussions, Sylvie Hénon for her help in handling the experiment, and Olivier Cardoso for assistance in bead tracking. This work was partly supported by a grant from "Association pour la Recherche sur le Cancer" (subvention libre # 3115). The biophysics group in MSC is also affiliated to the CNRS Consortium Cell Tiss (GDR 3070).

**NOTES AND <u>REFERENCES</u>**

§ In a few cases (8 cells), the bead diffusion was directly recorded with a fast camera (FASTCAM 1024/500) operating at 250 Hz, and tracked using a home-made ImageJ plugin. Their MSD did not present significant differences with the one measured with the quadrant photodiode, and all the data were gathered in the same pool.

1. Fabry B, *et al.* (2001) Scaling the microrheology of living cells. *Physical Review Letters* 8714(14):4.
2. Balland M, Richert A, & Gallet F (2005) The dissipative contribution of myosin II in the cytoskeleton dynamics of myoblasts. *European Biophysics Journal with Biophysics Letters* 34(3):255-261.
3. Fabry B, *et al.* (2003) Time scale and other invariants of integrative mechanical behavior in living cells. *Physical Review E* 68(4):18 .
4. Bausch AR, Moller W, & Sackmann E (1999) Measurement of local viscoelasticity and forces in living cells by magnetic tweezers. *Biophysical Journal* 76(1):573-579 (in English).
5. Wilhelm C, Gazeau F, & Bacri JC (2003) Rotational magnetic endosome microrheology: Viscoelastic architecture inside living cells. *Physical Review E* 67(6):12 .
6. Bursac P, *et al.* (2005) Cytoskeletal remodelling and slow dynamics in the living cell. *Nature Materials* 4(7):557-561 .
7. Desprat N, Richert A, Simeon J, & Asnacios A (2005) Creep function of a single living cell. *Biophysical Journal* 88(3):2224-2233 .
8. Lenormand G, Millet E, Fabry B, Butler JP, & Fredberg JJ (2004) Linearity and time-scale invariance of the creep function in living cells. *Journal of the Royal Society Interface* 1(1):91-97 .
9. Hoffman BD, Massiera G, Van Citters KM, & Crocker JC (2006) The consensus mechanics of cultured mammalian cells. *Proceedings of the National Academy of Sciences of the United States of America* 103(27):10259-10264 .
10. Balland M, *et al.* (2006) Power laws in microrheology experiments on living cells: Comparative analysis and modeling. *Physical Review E* 74(2):021911 .
11. Yamada S, Wirtz D, & Kuo SC (2000) Mechanics of living cells measured by laser tracking microrheology. *Biophysical Journal* 78(4):1736-1747.
12. Tseng Y, Kole TP, & Wirtz D (2002) Micromechanical mapping of live cells by multiple-particle-tracking microrheology. *Biophysical Journal* 83(6):3162-3176.

13. Lau AWC, Hoffman BD, Davies A, Crocker JC, & Lubensky TC (2003) Microrheology, stress fluctuations, and active behavior of living cells. *Physical Review Letters* 91(19).
14. Caspi A, Granek R, & Elbaum M (2002) Diffusion and directed motion in cellular transport. *Physical Review E* 66(1).
15. Weihs D, Mason TG, & Teitell MA (2006) Bio-microrheology: A frontier in microrheology. *Biophysical Journal* 91(11):4296-4305 .
16. Raupach C*, et al.* (2007) Stress fluctuations and motion of cytoskeletal-bound markers. *Physical Review E* 76(1).
17. Bursac P*, et al.* (2007) Cytoskeleton dynamics: Fluctuations within the network. *Biochemical and Biophysical Research Communications* 355(2):324-330.
18. Trepat X*, et al.* (2007) Universal physical responses to stretch in the living cell. *Nature* 447(7144):592-+.
19. Massiera G, Van Citters KM, Biancaniello PL, & Crocker JC (2007) Mechanics of single cells: Rheology, time depndence, and fluctuations. *Biophysical Journal* 93(10):3703-3713 .
20. Huet S*, et al.* (2006) Analysis of transient behavior in complex trajectories: Application to secretory vesicle dynamics. *Biophysical Journal* 91:3542-3559.
21. Mizuno D, Tardin C, Schmidt CF, & MacKintosh FC (2007) Nonequilibrium mechanics of active cytoskeletal networks. *Science* 315(5810):370-373.
22. Wilhelm C (2008) Out-of-equilibrium microrheology inside living cells. *Physical Review Letters* 101(2).
23. Mason TG & Weitz DA (1995) OPTICAL MEASUREMENTS OF FREQUENCY-DEPENDENT LINEAR VISCOELASTIC MODULI OF COMPLEX FLUIDS. *Physical Review Letters* 74(7):1250-1253.
24. Changeux JP, Pinset C, & Ribera AB (1986) EFFECTS OF CHLORPROMAZINE AND PHENCYCLIDINE ON MOUSE C2 ACETYLCHOLINE-RECEPTOR KINETICS. *Journal of Physiology-London* 378:497-513.
25. Icard-Arcizet D, Cardoso O, Richert A, & Henon S (2008) Cell stiffening in response to external stress is correlated to actin recruitment. *Biophysical Journal* 94(7):2906-2913.
26. Metzner C, Raupach C, Zitterbart DP, & Fabry B (2007) Simple model of cytoskeletal fluctuations. *Physical Review E* 76(2).
27. Abou B & Gallet F (2004) Probing a nonequilibrium Einstein relation in an aging colloidal glass. *Physical Review Letters* 93(16):4 .
28. Cugliandolo LF, Kurchan J, & Peliti L (1997) Energy flow, partial equilibration, and effective temperatures in systems with slow dynamics. *Physical Review E* 55(4):3898-3914.

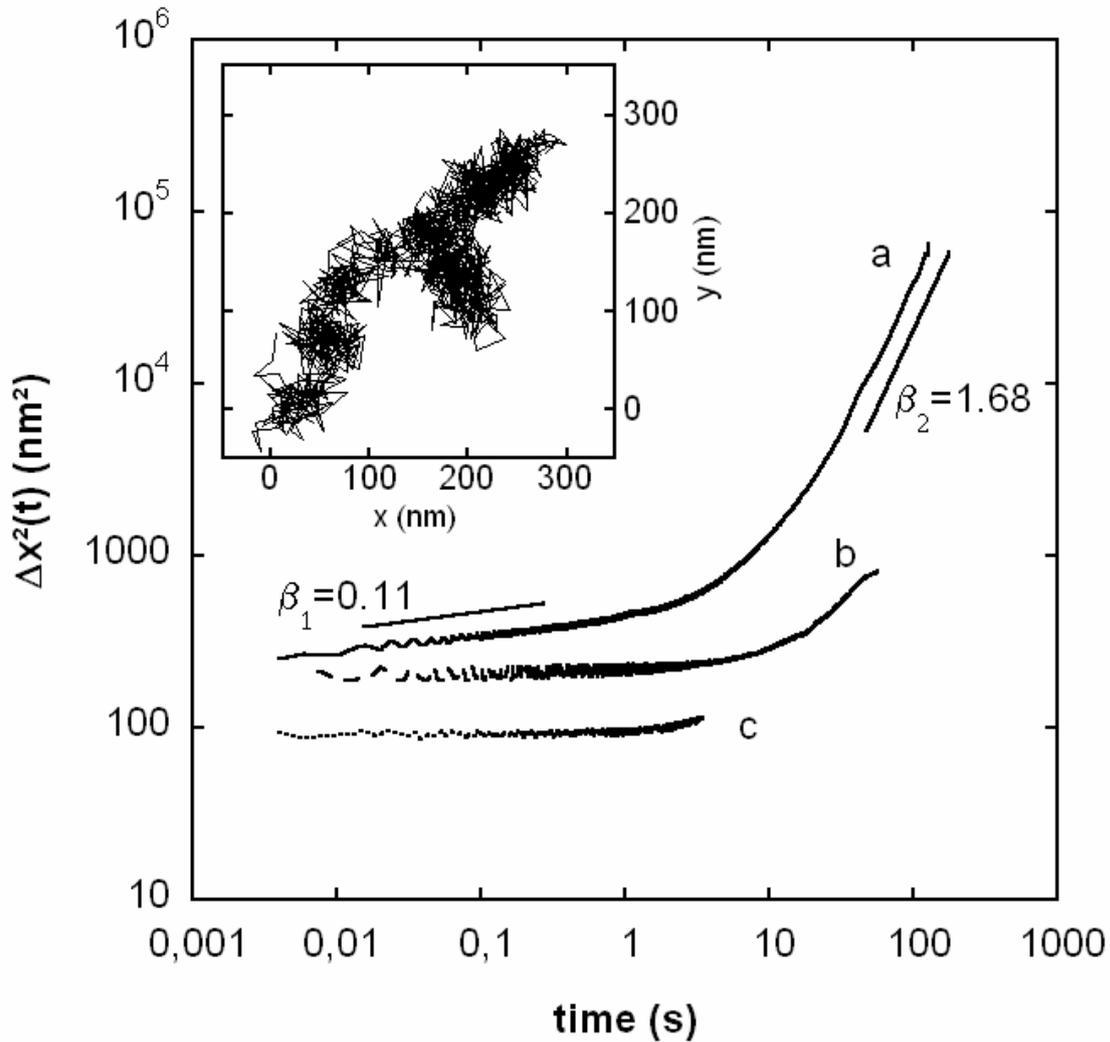

Figure 1:

Mean square displacement (MSD) $\Delta x^2(t)$ of a 1.56μm silica bead specifically bound, *via* the membrane, to the actin cortex of a C2-C12 cell: (a) In a control cell, the MSD exhibits two different regimes, each characterized by a power-law $t^\beta$ : subdiffusion at short time scales (t<3s, $\beta_1$ = 0.11), corresponding to a constrained motion of the bead, and superdiffusion at large time scales (t>3s, $\beta_2$ = 1.68), corresponding to a directed drift motion. (b) In an ATP-depleted cell, the MSD amplitude is lower and the bead motion remains subdiffusive even above t=10s. (c) Noise ground level, recorded from a bead stuck to the glass coverslip. Insert: bead trajectory in the x-y plane, associated to curve (a).

20/01/2009

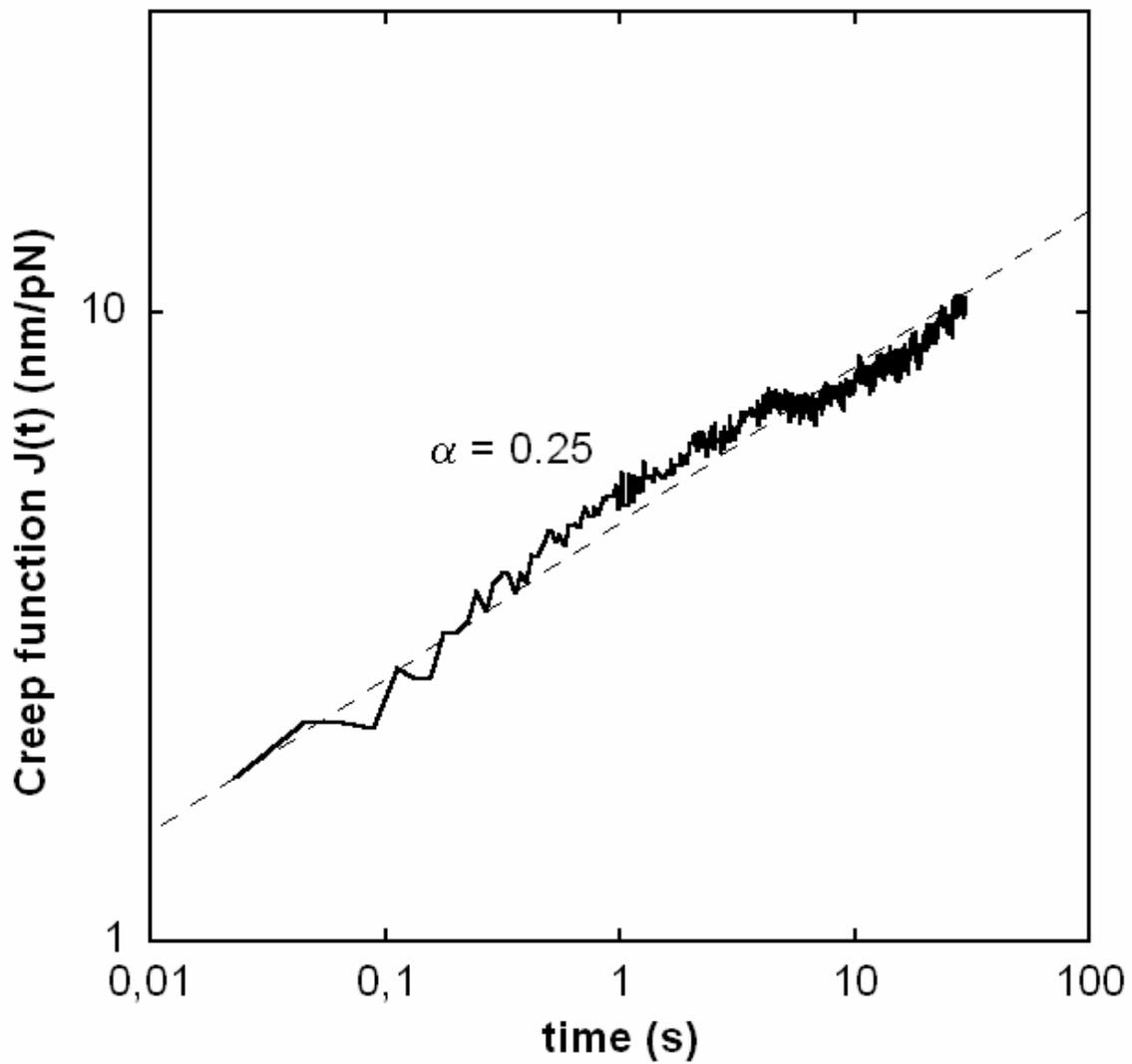

Figure 2:

Creep function J(t) of a bead bound to the cell membrane, in response to a force step. J(t) behaves as a weak power law over the full time range 0.02 - 30s. The best power law fit $J(t) = A(t/t_0)^\alpha$ leads to $\alpha = 0.25$. The average value over 60 measurements is $\langle\alpha\rangle = 0.180 \pm 0.015$.

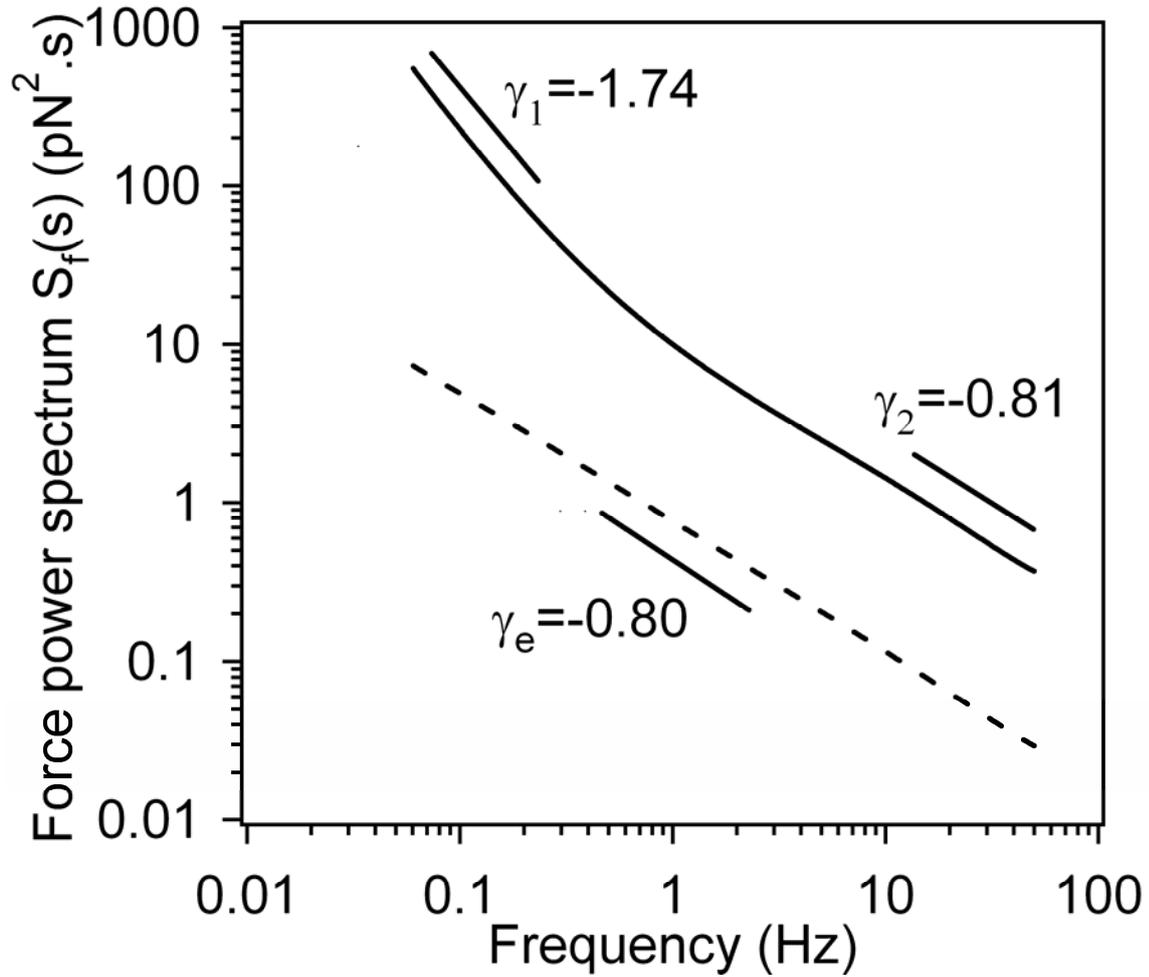

Figure 3:

Laplace Transform $\hat{S}_F(s)$ of the force-force correlation function, yielding the mean power spectrum of the fluctuating forces exerted on the probe (solid line: geometrical average over 60 measurements). Mean power spectrum $\hat{S}_{Feq}(s)$, calculated as if the system was at equilibrium (dashed line). We distinguish two different regimes, expressed as frequency power laws $\hat{S}_F(s) \propto s^{-\gamma}$. At high frequencies s>0.5 Hz the exponent $\gamma_2$ = -0.81 is close to the exponent $\gamma_e$ = -0.80 retrieved from $\hat{S}_{Feq}(s)$, but the ratio $\hat{S}_F(s)/\hat{S}_{Feq}(s) \approx$ 13. At lower frequencies s<0.5 Hz, the exponent is shifted to $\gamma_1$= -1.74, and the ratio $\hat{S}_F(s)/\hat{S}_{Feq}(s)$ rapidly increases with decreasing s. The crossover at s ≈ 0.5 Hz (t ≈ 2s) corresponds to the transition between the subdiffusive and the superdiffusive regimes observed during free particle motion.

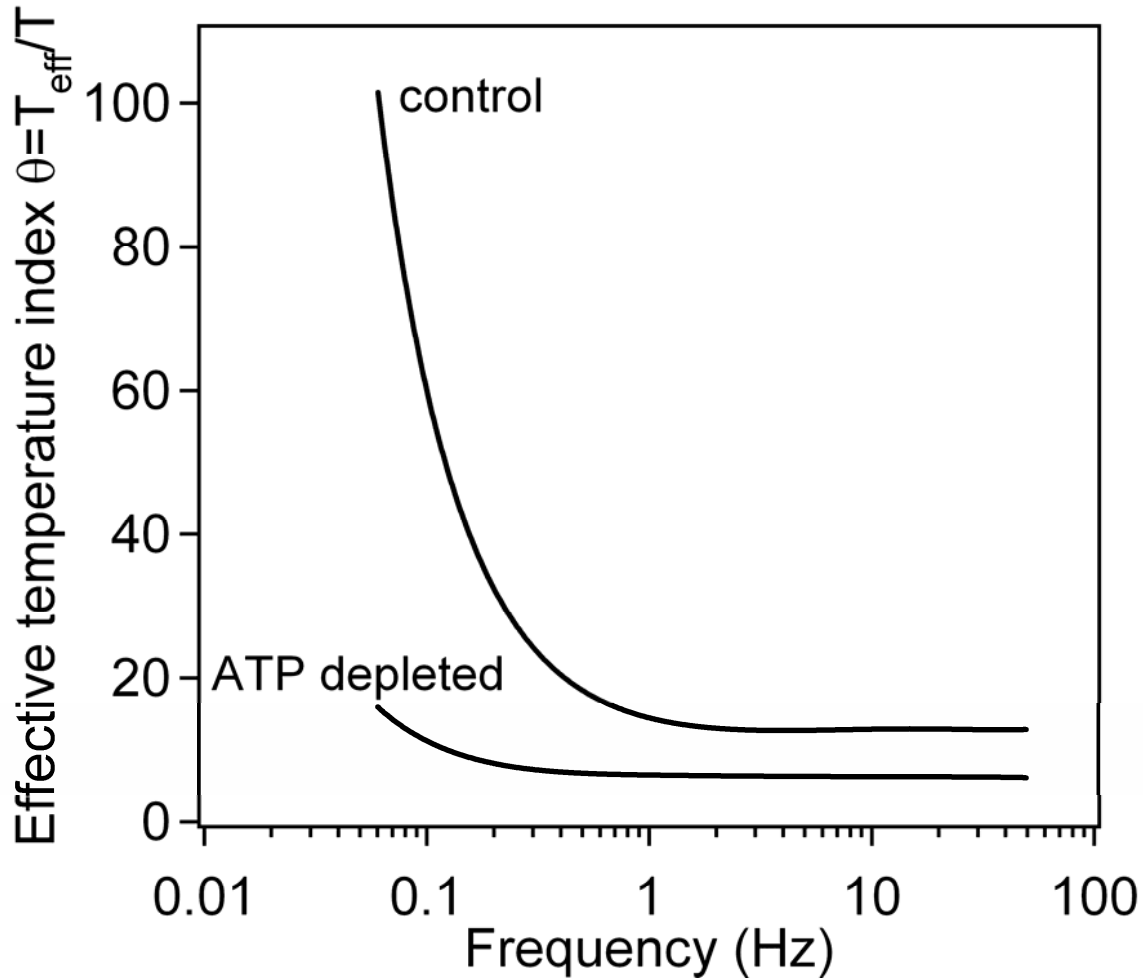

Figure 4:

Plots of $\theta(s) = T_{eff}/T$, ratio of the effective temperature of the system to the bath temperature, versus frequency. Upper curve: control cells (60 measurements); lower curve: ATP-depleted cells (23 measurements). In control conditions, $\theta(s)$ remains constant, close to 13, for $s > 0.5$ Hz, and rapidly increases as s decreases below 0.5 Hz. For ATP-deprived cells, $\theta(s)$ is substantially lowered, indicating that ATP depletion considerably reduces the active part of force fluctuations and brings the cells back closer to equilibrium conditions. At high frequency, $\theta(s)$ is constant, around 6 (2 times smaller than in control cells); at lower frequency, although $\theta(s)$ slightly increases with decreasing s, it is reduced by a factor up to 10 as compared to controlled conditions.